\documentclass[preprint,superscriptaddress,showkeys,showpacs,tightenlines,fleqn,prd]{revtex4} 
\usepackage{graphicx}
\begin{document}      
\preprint{PNU-NTG-05/2007}
\preprint{PNU-NuRI-05/2007}
\title{ Kaon semileptonic decay ($K_{l3}$) form factor\\ 
in the nonlocal chiral quark model}      
\author{Seung-il Nam}
\email{sinam@yukawa.kyoto-u.ac.jp, sinam@pusan.ac.kr}
\affiliation{Yukawa Institute for Theoretical Physics (YITP), Kyoto
University, Kyoto 606-8502, Japan} 
\affiliation{Department of
Physics and Nuclear Physics \& Radiation Technology Institute (NuRI),
Pusan National University, Busan 609-735, Republic of Korea} 
\author{Hyun-Chul Kim}
\email{hchkim@pusan.ac.kr}
\affiliation{Department of
Physics and Nuclear Physics \& Radiation Technology Institute (NuRI),
Pusan National University, Busan 609-735, Republic of Korea} 
\date{\today}
\begin{abstract}   
We investigate the kaon semileptonic decay ($K_{l3}$) form 
factors within the framework of the nonlocal chiral quark model 
($\chi$QM) from the instanton vacuum, taking into account the effects 
of flavor SU(3) symmetry breaking. All theoretical calculations are
carried out without any adjustable parameter.  We also show that the 
present results satisfy the Callan-Treiman low-energy theorem as well 
as the Ademollo-Gatto theorem.  It turns out that 
the effects of flavor SU(3) symmetry breaking are essential in 
reproducing the kaon semileptonic form factors.  The present results 
are in a good agreement with experiments, and are compatible with 
other model calculations.
\end{abstract} 
\maketitle
\section{Introduction}
It is of great importance to understand semileptonic decays of kaons 
($K_{l3}$), since it plays a significant role in determining the CKM 
matrix element $|V_{us}|$ 
precisely~\cite{Cabibbo:1963yz}. The effect of flavor
SU(3) symmetry breaking on the kaon semileptonic  
decay form factor is known to be around $3\sim5\%$, which is rather 
small. The well-known soft-pion Callan-Treiman~\cite{Callan:1966hu} theorem 
connects the ratio of the pion and kaon decay constants to the 
semileptonic form factors of the kaon at $q^2=m_K^2-m_\pi^2$ 
(Callan-Treiman point). Experimentally, there are a certain amount of
data to judge  
theoretical calculations~\cite{Yao:2006px}. Thus, the 
kaon semileptonic decay form factor provides a basis to examine the  
validity and reliability of any theoretical theory and model for 
hadrons. Related works on the kaon semileptonic decay can be found in Refs.~\cite{Gershtein:1976mv,Khlopov:1978id,Gasser:1984ux,Leutwyler:1984je,Bijnens:1994me,Becirevic:2004ya,Tsutsui:2005cj,Ji:2001pj,Kalinovsky:1996ii,Isgur:1975hu,Choi:1998jd,Choi:1999nu}, 

In the present work, we will investigate the $K_{l3}$ form factor 
within the framework of the nonlocal chiral quark model ($\chi$QM) 
derived from the instanton vacuum.  We will consider the leading order 
in the large $N_c$ expansion and flavor SU(3) symmetry breaking 
explicitly.  The model has several virtues: All relevant QCD 
symmetries are satisfied within the model, and there are only two: 
The average size of instantons ($\rho\sim1/3$ fm) and average  
inter-instanton distance ($R\sim1$ fm), which can be determined by the  
internal constraint such as the self-consistent 
equation~\cite{Shuryak:1981ff,Diakonov:1983hh}. There
is no further adjustable parameter in the model. 

We employ the  
modified low-energy effective partition function with flavor SU(3) 
symmetry 
breaking~\cite{Musakhanov:2001pc}.   
This partition function extends the former one derived in the chiral 
limit~\cite{Diakonov:1983hh}.  It has been proven that 
the partition function with flavor SU(3) symmetry breaking is very 
successful in describing the low-energy hadronic properties such as 
various QCD condensates, magnetic susceptibilities, meson distribution 
amplitudes, and so 
on~\cite{Kim:2004hd,Nam:2006sx,Nam:2006ng}.  
However, the presence of the nonlocal interaction between quarks and 
pseudo-Goldstone bosons breaks the Ward-Takahashi identity for the 
N\"other currents.  Since the kaon semileptonic decay form factors 
involve the vector current, we need to deal with this problem. Thus, in the 
present work, we will investigate the kaon semileptonic decay 
($K_{l3}$) form factors, using the gauged low-energy effective  
partition function from the instanton vacuum with flavor SU(3) 
symmetry breaking explicitly taken into account. 
\section{Formalisms}
In the present work, we are interested in the following kaon semileptonic 
decays ($K_{l3}$) in two different isospin channels:  
\begin{eqnarray} 
  \label{eq:DEF1} 
K^+(p_K)&\to&\pi^0(p_{\pi})\,l^+(p_l)\,\nu_l(p_{\nu}):\,K^{+}_{l3}, 
\nonumber\\ 
K^0(p_K)&\to&\pi^-(p_{\pi})\,l^+(p_l)\,\nu_l(p_{\nu}):\,K^{0}_{l3}, 
\end{eqnarray} 
where $l$ and $\nu_l$ stand for the leptons (either the electron or 
the muon) and neutrinos. The decay amplitude ($T_{K\to l\nu\pi}$) can
be expressed as  
follows~\cite{Bijnens:1994me}:   
\begin{equation} 
  \label{eq:T} 
T_{K\to l\nu\pi}=\frac{G_F}{\sqrt{2}}\sin\theta_c\left[w^{\mu}(p_l,p_{\nu}) 
F{\mu}(p_K,p_{\pi})\right],  
\end{equation} 
where $G_F$ is the well-known Fermi constant 
($1.166\times10^{-5}\,{\rm{GeV}}^{-2}$). $\theta_c$ denotes the   
Cabbibo angle. We define respectively the weak leptonic current 
($w^{\mu}$) and hadronic matrix element $F_{\mu}$ with the $\Delta 
S=1$ vector current ($j^{su}_{\mu}$) as:    
\begin{eqnarray} 
  \label{eq:DEF2} 
w^{\mu}(p_l,p_{\nu})&=&\bar{u}(p_{\nu})\gamma^{\mu}(1-\gamma_5)v(p_l), 
\\ 
\label{eq:DEF3} 
F_{\mu}(p_K,p_{\pi}) 
&=&c\langle \pi(p_\pi)|j^{su}_{\mu}|K(p_K)\rangle 
=c\langle \pi(p_\pi)|\bar{\psi}\gamma_{\mu}\lambda^{4+i5} 
\psi|K(p_K)\rangle  
\nonumber\\ 
&=&(p_K+p_{\pi})_{\mu}f_{l+}(t)+(p_K-p_{\pi})_{\mu}f_{l-}(t), 
\end{eqnarray}  
where $c$ is the isospin factor, and set to be unity and $1/\sqrt{2}$ 
for $K^0_{l3}$ and $K^+_{l3}$, respectively.  The matrix 
$\lambda^{4+i5}$ denotes the combination of the two Gell-Mann  
matrices, $\left(\lambda^4+i\lambda^5\right)/2$, for the 
relevant flavor in the present problem.  The $\psi$ denotes the quark   
field.  The momentum transfer is defined as 
$Q^2=(p_K-p_{\pi})^2\equiv{-t}$.    
 
$f_{l\pm}$ represent the vector form factors with the corresponding 
lepton $l$ ($P$-wave projection).  Alternatively, the form factor 
$F_{\mu}(p_K,p_{\pi})$ can be expressed in terms of the scalar 
($f_{l0}$, $S$-wave projection) and the vector form factor $f_{l+}$ 
defined as follows:   
\begin{equation} 
  \label{eq:scalar} 
F_{\mu}(p_K,p_{\pi})=f_{l+}(t)(p_K+p_{\pi})_{\mu} 
+\frac{(m^2_{\pi}-m^2_K)(p_K-p_{\pi})_{\mu}}{t} 
\left[f_{l+}(t)-f_{l0}(t)\right].   
\end{equation} 
Hence, the $f_{l0}$ can be written as the linear combination of 
$f_{l+}$ and $f_{l-}$:   
\begin{equation} 
  \label{eq:swave}  
  f_{l0}(t)=f_{l+}(t)+\left[\frac{t}{m^2_K-m^2_{\pi}}\right]f_{l-}(t). 
\end{equation} 
Since the isospin breaking effects are almost negligible, we 
will consider only the $K^0\to\pi^-\nu{l}^+$ decay channel. 

It has been well-known that the experimental data for $f_{l+,0}$ can 
be reproduced qualitatively well by the linear and quadratic 
fits~\cite{Yao:2006px}:     
\begin{eqnarray} 
  \label{eq:ex} 
{\rm Linear}&:&\,f_{l+,0}(t)=f_{l+,0}(0) 
\left[1+\frac{\lambda_{l+,0}}{m^2_{\pi}}(t-m^2_l) 
\right], 
\nonumber\\ 
{\rm Quadratic}&:&\,f_{l+,0}(t) 
=f_{l+,0}(0)\left[1+\frac{\lambda'_{l+,0}}{m^2_{\pi}}(t-m^2_l) 
+\frac{\lambda''_{l+,0}}{2m^4_{\pi}}(t-m^2_l)^2\right], 
\end{eqnarray} 
where $m_l$ is the lepton mass. The slope parameter $\lambda_{l+}$ has 
an important physical meaning.  For example, the $K\to\pi$ decay radius    
($\langle r^2\rangle^{K\pi}$) can be obtained as  
follows~\cite{Bijnens:1994me}:  
\begin{equation} 
  \label{eq:cr} 
\lambda_+\simeq\frac{1}{6}\langle{r^2}\rangle^{K\pi}m^2_{\pi}. 
\end{equation} 
Moreover, this radius is also related to the Gasser-Leutwyler 
low-energy constant $L_9$ in the large $N_c$ 
limit~\cite{Gasser:1984ux} as follows:   
\begin{equation} 
  \label{eq:gasser} 
L_9=\frac{1}{12}F^2_{\pi}\langle r^2 \rangle^{K\pi}. 
\end{equation} 

We now show how to derive the hadronic matrix element 
given in Eq.~(\ref{eq:DEF3}) within the framework of the nonlocal 
$\chi$QM from the instanton vacuum.  We begin by the low-energy  
effective QCD partition function derived from the instanton 
vacuum~\cite{Musakhanov:2001pc}:  
\begin{eqnarray}   
{\cal Z}_{\rm eff.}&=&\int{\cal D}\psi{\cal D}\psi^\dagger  
{\cal D} {\cal M} \exp\int d^4 x\Big[ \psi^{\dagger}_{f} (x)   
(i\rlap{/}{\partial }+im_f)\psi_{f}(x)\nonumber   
\\&+&i\int\frac{d^4k\,d^4p}{(2\pi)^8} e^{-i(k-p)\cdot x}  
\psi^{\dagger}_{f}(k)\sqrt{M_f(k_{\mu})}  
U^{\gamma_5}_{fg} \sqrt{M_g(p_{\mu})}\psi_{g}(p)\Big].    
\label{effectiveaction} 
\end{eqnarray}  
$M_f(k)$ is 
the dynamically generated quark mass being momentum-dependent, whereas 
$m_f$ stands for the current-quark mass with flavor $f$. $U^{\gamma
_{5}}$ is the nonlinear 
background Goldstone boson field. As mentioned
previously, the momentum-dependent dynamical quark mass  
$M_f(k)$ breaks the conservation of the N\"other (vector) currents. 
Refs.~\cite{Kim:2004hd} derived  
the light-quark partition function in the presence of the external  
vector field. By doing so, we can derive the gauge-invariant formula
for the kaon semileptonic form factor as follows:
\begin{eqnarray} 
  \label{eq:local} 
F^{\rm local(a)}_{\mu}&=&\frac{8N_c}{F_{\pi}F_K}\int\frac{d^4k}{(2\pi)^4} 
\frac{M_q(k_a)\sqrt{M_s(k_b)M_q(k_c)}} 
{\left[k^2_a+\bar{M}^2_q(k_a)\right] 
\left[k^2_b+\bar{M}^2_s(k_b)\right] 
\left[k^2_c+\bar{M}^2_q(k_c)\right]} 
\nonumber\\ 
&\times& 
\Bigg[\left[k_a\cdot{k_b}+\bar{M}_q(k_a)\bar{M}_s(k_b)\right] 
k_{c\mu}-\left[k_b\cdot{k_c}+\bar{M}_s(k_b)\bar{M}_q(k_c)\right]k_{a\mu} 
\nonumber\\ 
&+& 
\left[k_a\cdot{k_c}+\bar{M}_q(k_a)\bar{M}_q(k_c)\right]k_{b\mu}\Bigg], 
\end{eqnarray} 
where $\bar{M}_f(k)=m_f+M_f(k)$. The relevant momenta are defined as 
$k_a=k-p/2-q/2$, $k_b=k+p/2-q/2$ and $k_c=k+p/2+q/2$, in which $k$, 
$p$ and $q$ denote the internal quark, initial kaon, and 
transfered momenta, respectively.  The trace ${\rm tr}_{\gamma}$ runs  
over Dirac spin space.  Similarly, we can evaluate the nonlocal 
contributions as follows~\cite{Nam:2006sx}:    
\begin{eqnarray}  
\label{eq:nonlocal} 
&&F^{\rm nonlocal(b)}_{\mu} 
=\frac{8N_c}{F_{\pi}F_K}\int\frac{d^4k}{(2\pi)^4} 
\frac{\sqrt{M_q(k_c)}_{\mu}\sqrt{M_q(k_c)}M_q(k_a)M_s(k_b)} 
{\left[k^2_a+\bar{M}^2_q(k_a)\right] 
\left[k^2_b+\bar{M}^2_s(k_b)\right] 
\left[k^2_c+\bar{M}^2_q(k_c)\right]} 
\nonumber\\ 
&&\times 
\left[\bar{M}_q(k_c)k_a\cdot k_b+\bar{M}_s(k_b)k_a\cdot 
  k_c-\bar{M}_q(k_a)k_b\cdot k_c 
+\bar{M}_q(k_a)\bar{M}_s(k_b)\bar{M}_q(k_c) \right] 
\nonumber\\
&&-\left(b\leftrightarrow c\right), 
\nonumber\\ 
&&F^{\rm nonlocal(c)}_{\mu} 
=-\frac{4N_c}{F_{\pi}F_K}\int\frac{d^4k}{(2\pi)^4} 
\nonumber\\
&&\times
\frac{ 
\sqrt{M_q(k_a)}\sqrt{M_s(k_b)}\sqrt{M_q(k_c)}_{\mu}\sqrt{M_q(k_a)} 
\left[k_a\cdot k_b+\bar{M}_q(k_a)\bar{M}_s(k_b)\right]} 
{\left[k^2_a+\bar{M}^2_q(k_a)\right]\left[k^2_b+\bar{M}^2_s(k_b)\right]} 
\nonumber\\ 
&&+\frac{4N_c}{F_{\pi}F_K}\int\frac{d^4k}{(2\pi)^4} 
\nonumber\\
&&\times
\frac{\sqrt{M_q(k_a)}\sqrt{M_s(k_b)}\sqrt{M_q(k_c)}\sqrt{M_q(k_a)}_{\mu} 
\left[k_a\cdot k_b+\bar{M}_q(k_a)\bar{M}_s(k_b)\right]} 
{\left[k^2_a+\bar{M}^2_q(k_a)\right]\left[k^2_b+\bar{M}^2_s(k_b)\right]} 
\nonumber\\ 
&&+\left(b\leftrightarrow c\right), 
\end{eqnarray} 
where $\sqrt{M(k)}_{\mu}=\partial\sqrt{M(k)}/\partial k_{\mu}$. 
\section{Numerical results}
\begin{figure}[t] 
\begin{center} 
\begin{tabular}{cc} 
\includegraphics[width=8cm]{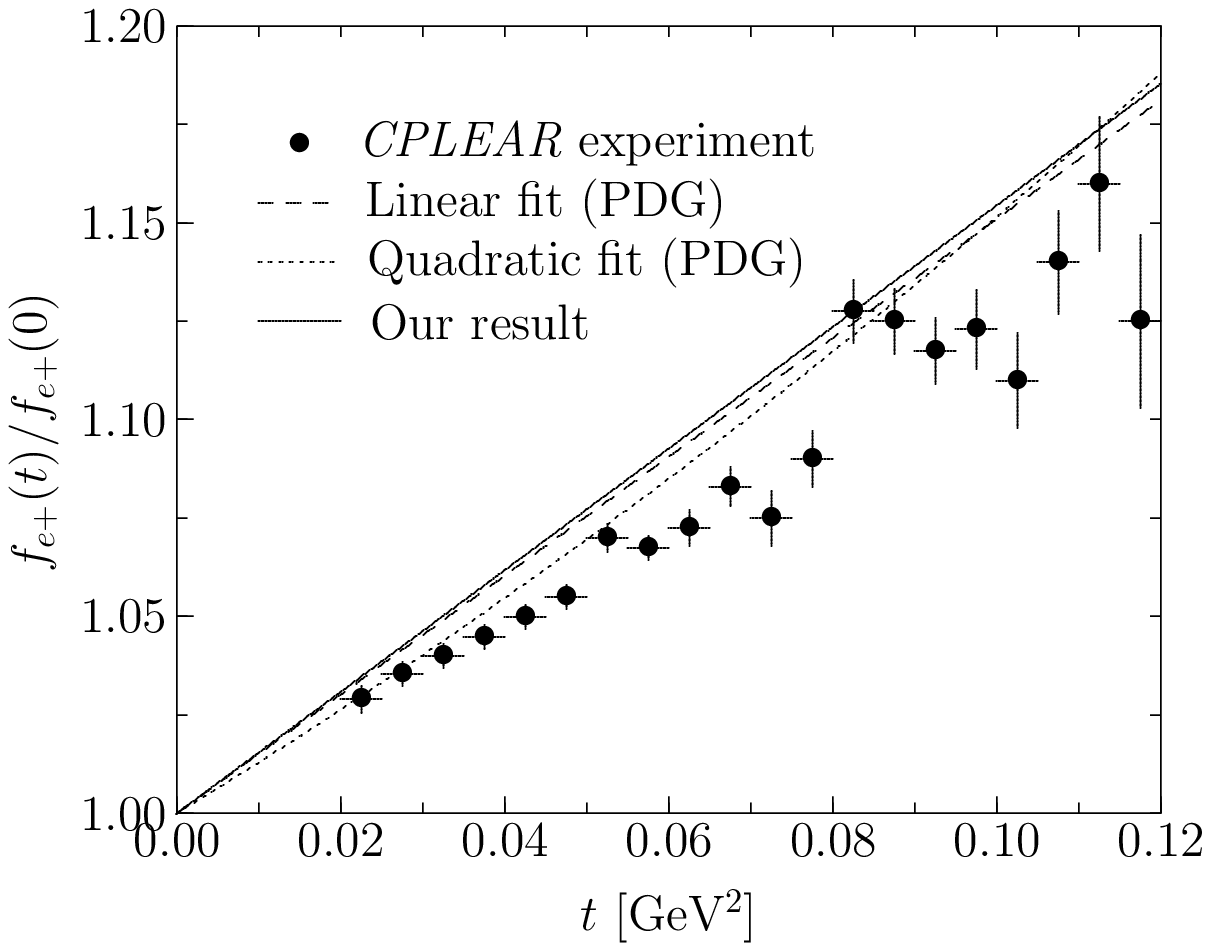} 
\includegraphics[width=8cm]{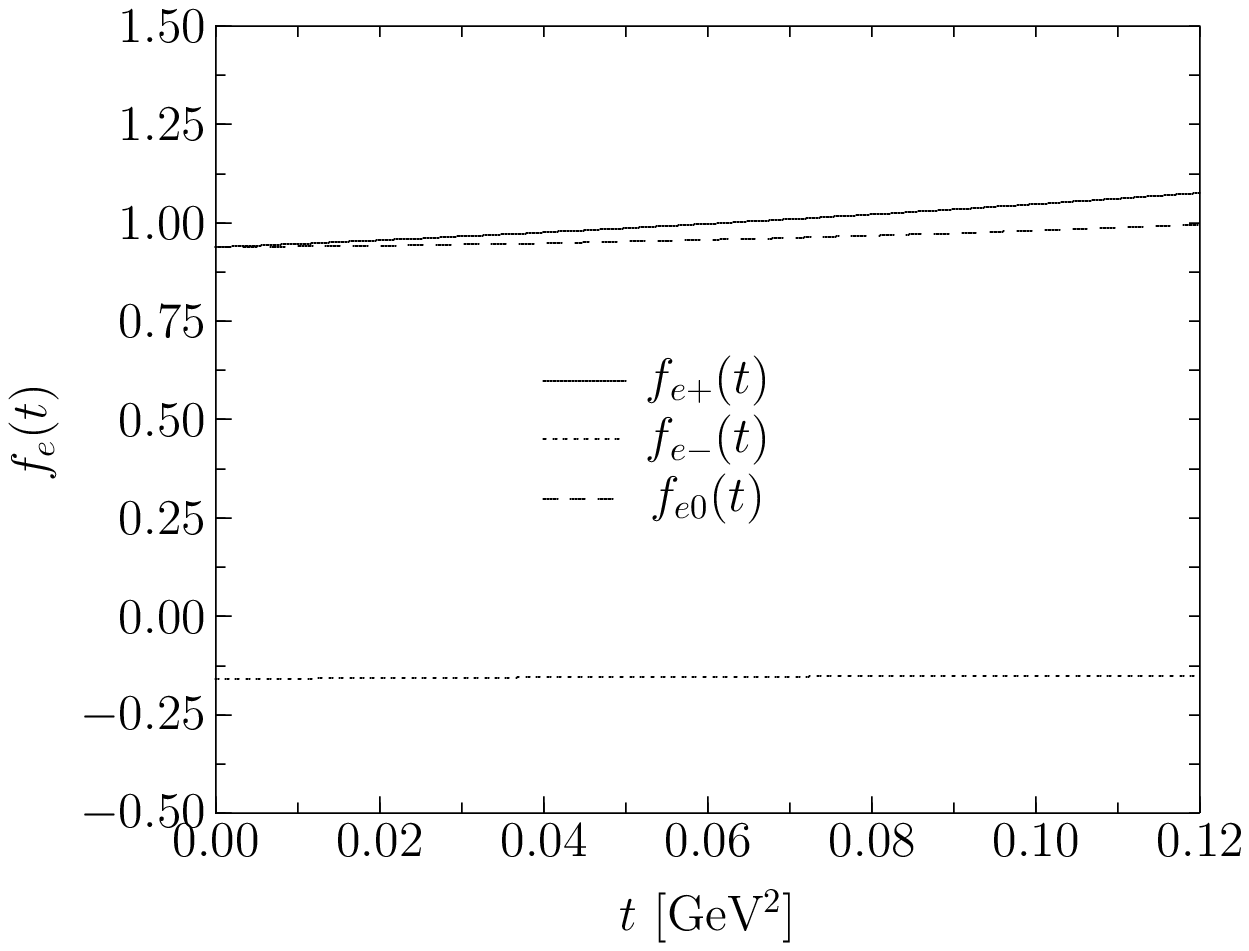} 
\end{tabular} 
\end{center} 
\caption{$K_{e3}$ form factors, $f_{e+}(t)$ (solid), $f_{e-}(t)$ (dotted) and 
$f_{e0}(t)$ (dashed) are shown in   the left panel, while in the right  
panel the ratio of $f_{e+}(t)$ and $f_{e+}(0)$ is given (solid).}          
\label{fig1}
\end{figure}

We now discuss various numerical results for the kaon semileptonic 
decay ($K_{l3}$) form factors in the present work.  We facilitate the  
Breit-momentum framework for convenience by virtue of the Lorentz
invariance of the model. We first consider the case of $K_{e3}$. In
the left panel of Figure~\ref{fig1}, we draw the  
numerical results for $f_{e+}(t)$ (solid), $f_{e-}(t)$ (dotted) and 
$f_{e0}(t)$ (dashed).  Note that the scalar form factor 
$f_{e0}(t)$ is derived by using Eq.~(\ref{eq:swave}).   
We observe that the $f_{e+}(t)$ and $f_{e0}(t)$ are almost linearly 
increasing functions of $t$, whereas $f_{e-}(t)$ decreases. At $t=0$, our   
results demonstrate that $f_{e+}(0)=f_{e0}(0)=0.947$ and 
$f_{e-}(0)=-0.137$.  In the chiral limit, $f_{e+}(0)$ and $f_{e-}(0)$ 
should be unity and zero, respectively, which is related to the     
Ademollo-Gatto theorem in the case of pseudo-Goldstone 
bosons~\cite{Gasser:1984ux,Ademollo:1964sr,Langacker:1973nf}:   
\begin{equation} 
\label{eq:GA} 
\lim_{q\to0}F^{\rm local(a)}_{\mu}\simeq 2p_{\mu} + \mathcal{O}(m_q). 
\end{equation} 
The Ademollo-Gatto theorem in Eq.~(\ref{eq:GA}) can be easily tested in the
nonlocal $\chi$QM. Considering $q\to0$ and ignoring the terms being  
proportional to $k\cdot p$, the leading contribution of 
Eq.~(\ref{eq:local}) can be rewritten upto ${\cal O}(m_q)$ as follows:  
\begin{equation}
  \label{eq:AG}
\lim_{q\to0}F^{\rm local(a)}_{\mu}\simeq2\left[1+R(m_s)\right]p_{\mu},
\end{equation}
where
\begin{equation}
  \label{eq:AG1}
R(m_s)=\frac{1}{2}\left[\int\frac{d^4k}{(2\pi)^4} 
\frac{M^2(k)m_s\left[m_s+2M(k)\right]}
{\left[k^2+M^2(k)\right]^3}\right]
\left[\int\frac{d^4k}{(2\pi)^4} 
\frac{M^2(k)}
{\left[k^2+M^2(k)\right]^2}\right]^{-1}.  
\end{equation}
To evaluate Eq.~(\ref{eq:AG}), we employ the ratio $F_K/F_{\pi}$
computed within the same framework and expanded in terms of
the strange quark mass ($m_s$):   
\begin{equation}
  \label{eq:PS}
\frac{F_K}{F_{\pi}}\simeq1+R(m_s).
\end{equation} 
We also use that $k_b=k_c\to k+p/4$ since these
two momenta share $p/2$ as  
$q\to0$. Note that we consider only the local
contribution for $F_{\cal M}$ in  
Eq.~(\ref{eq:PS}). We, however, verified that  
the nonlocal contributions in  
Eq.~(\ref{eq:nonlocal}) also satisfies the Ademollo-Gatto theorem 
analytically.  
 
In the right panel of Figure~\ref{fig1} we draw the ratio of 
$f_{e+}(t)$ and $f_{e+}(0)$ with respect to the CPLEAR experimental 
data~\cite{Apostolakis:1999gs}, and linear (dashed)  and quadratic (dotted) 
fits for the ratio using the PDG data~\cite{Yao:2006px}: 
$\lambda_{e+}=(2.960\pm0.05)\times10^{-2}$, 
$\lambda'_{e+}=(2.485\pm0.163)\times10^{-2}$, and    
$\lambda''_{e+}=(1.920\pm0.062)\times10^{-3}$.  In the present calculation,  
we obtain $\lambda_{e+}=3.028\times10^{-2}$ for the linear fit, which 
is very close to the experimental one, $2.960\times10^{-2}$.  Since 
our result for $f_{e+}$ is almost linear as shown in Fig.~\ref{fig1}, 
we get almost a negligible value for the slope parameter $\lambda''$ 
when the quadratic fit is taken into account. Being compared with  
other model calculations, the present results are comparable 
to those from $\chi$PT~\cite{Bijnens:1994me}, and 
other
models~\cite{Ji:2001pj,Kalinovsky:1996ii,Afanasev:1996my,Tsutsui:2005cj}.   
Using Eq.~(\ref{eq:cr}) and 
Eq.~(\ref{eq:gasser}), we can easily  
estimate the $K_{e3}$ decay radius and low-energy constant $L_9$,   
respectively.  As for the $K_{e3}$ decay radius, we obtain  
$\langle{r}^2\rangle^{K\pi}=0.366\,{\rm{fm}}^2$.  This value is 
slightly larger than that in $\chi$PT~\cite{Gasser:1984ux}. The
low-energy constant $L_9$ turns out   
to be $6.78\times10^{-3}$, which is comparable to 
$7.1\sim7.4\times10^{-3}$~\cite{Gasser:1984ux} and  
$6.9\times10^{-3}$~\cite{Bijnens:1994me,Ecker:1994gg}.     
 
The ratio of the pion and kaon weak decay constants  
$F_K/F_{\pi}$ can be deduced from the scalar form factor $f_0$ via the 
Callan-Treiman soft-pion theorem~\cite{Callan:1966hu}. In the 
soft-pion limit ($p_{\pi}\to0$), the $K_{e3}$ form factor can be 
written as~\cite{lee:1968,Passemar:2006tc,Werth:2006tj}:  
\begin{equation} 
\label{eq:CT2} 
\lim_{p_{\pi}\to0}F_{\mu}(p_{\pi},p_K)=p_{K\mu}\frac{F_K}{F_\pi}.   
\end{equation} 
Using Eqs.~(\ref{eq:DEF3}) and (\ref{eq:swave}), we obtain the 
following expression: 
\begin{equation} 
  \label{eq:CT1} 
\lim_{p_{\pi}\to0}F_{\mu}(p_{\pi},p_K)= 
\lim_{p_{\pi}\to0}(p_{\pi}+p_K)_{\mu} 
\left[f_{l+}(\Delta_{\rm CT})+f_{l-}(\Delta_{\rm CT})\right] 
\simeq p_{K\mu}f_{l0}(\Delta_{\rm CT}), 
\end{equation} 
where the value of $\Delta_{\rm CT}=m^2_K-m^2_{\pi}$ is called the 
Callan-Treiman point which can not be accessible physically.  Combining 
Eq.~(\ref{eq:CT2}) with Eq.(\ref{eq:CT1}), we finally arrive at 
the final expression of the $K_{l3}$ form factor for the 
Callan-Treiman theorem in terms of the scalar form factor and the 
ratio, $F_K/F_{\pi}$:    
\begin{equation} 
  \label{eq:CT} 
f_{e0}(m^2_K-m^2_{\pi})=\frac{F_K}{F_{\pi}}. 
\end{equation} 
From our numerical calculation using Eq.~(\ref{eq:CT}) we find that 
$F_K/F_{\pi}=1.08$, which is $\sim10\%$ smaller than the empirical 
value ($1.22$). This smallness is mainly depends on the nonlocal 
contributions (c) in Eq.~(\ref{eq:nonlocal}) such that $f_{e-}$ decrease as 
depicted in the left panel of Figure~\ref{fig1}. This behavior can be  
interpreted by the fact that the kaon weak decay constant turns out to 
be smaller if we ignore the meson-loop correction in the nonlocal 
$\chi$QM~\cite{Kim:2005jc} and in chiral  
perturbation theory ($\chi$PT) as well, in which the ratio is defined  in the 
large $N_c$ limit by: 
\begin{equation} 
  \label{eq:chiral} 
\frac{F_K}{F_{\pi}}=1+\frac{4}{F^2_{\pi}}\left(m^2_K-m^2_{\pi}\right)L_5.   
\end{equation} 
Using the value of $F_K/F_{\pi}=1.08$, we obtain 
$L_5=7.67\times10^{-4}$ which is underestimated by a half of the 
phenomenological one $1.4\times10^{-3}$~\cite{Ecker:1994gg}. It is 
well known that in order to reproduce the $L_5$ within the $\chi$QM  
the meson-loop $1/N_c$ corrections are essential.   
    
In the soft limit, the model should satisfy the Callan-Treiman 
theorem given in Eq.~(\ref{eq:CT}).  Taking the limit $p_{\pi}\to0$ 
for Eq.~(\ref{eq:local}), we can show that Eq.~(\ref{eq:local})  
satisfies the Callan-Treiman theorem using Eq.~(\ref{eq:PS}) as follows: 
\begin{equation} 
  \label{eq:let} 
\lim_{p_{\pi}\to0} F^{\rm local(a)}_{\mu}\simeq\left[1+R(m_s)\right]p_{\mu},  
\end{equation} 
where $k_a= k_c\to k$ as $p_{\pi}\to 0$.  Inserting Eq.~(\ref{eq:PS}) 
into Eq.~(\ref{eq:let}), we can verify the validity of the 
Callan-Treiman theorem in Eq.~(\ref{eq:CT2}) (Eq.~(\ref{eq:CT})). The 
same argument also holds for the nonlocal contributions.  
 
The decay width of  $K\to\pi\nu{e}$ can be easily computed by using 
the result of $f_{l+,0}$.  It turns out that  
$\Gamma_{e3}=6.840\times10^6/$s and $\Gamma_{\mu3}=4.469\times10^6/$s 
with $|V_{us}|=0.22$ taken into 
account~\cite{Yao:2006px,Calderon:2001ni}. The results are slightly 
smaller than the experimental data ($\Gamma_{e3}=(7.920\pm0.040)\times10^6/$s 
and $\Gamma_{\mu3}=(5.285\pm0.024)\times10^6/$s)~\cite{Yao:2006px}.   

\section{Summary and conclusion}
In the present work, we have investigated the kaon semileptonic decay 
($K_{l3}$) form factors within the framework of the gauged nonlocal 
chiral quark model from the instanton vacuum.  The effect of flavor 
SU(3) symmetry breaking were taken into account.  We calculated the 
vector form factors ($f_{\pm}$), scalar form factor ($f_0$),  
slope parameters ($\lambda_{+,0}$), decay width ($\Gamma_{l3}$), etc.
We found that the present results  
of the kaon semileptonic decay form factors are in a qualitatively good 
agreement with experiments. We emphasize that there were no adjustable free 
parameters in the present investigation.  All results were obtained 
with only two parameters from the instanton vacuum, i.e. the average 
instanton size ($\bar{\rho}\sim1/3$ fm) and inter-instanton distance 
($R\sim1$ fm).  
 
In the present investigation, we have considered only the 
leading-order contributions in the large $N_c$ limit.  While these 
contributions reproduce the observables relevant for kaon semileptonic 
decay in general, it seems necessary to take into account the $1/N_c$ 
meson-loop corrections in order reproduce quantitatively the kaon 
decay constant $f_K$ and the low-energy constant $L_5$.  As noticed in   
Refs.~\cite{Nam:2006au,Gasser:1984gg,Kim:2005jc}, this correction for the  
fluctuation (meson-loop correction) can play an important role in producing 
the kaon properties as shown in the ratio $F_K/F_{\pi}$ as discussed and 
showed in the text. Moreover, it was shown that some of the low energy 
constants are very sensitive to this correction.  Related works are 
under progress. For more details on the present work, one can refer to
Ref.~\cite{Nam:2007fx}. 
\section*{Acknowledgements}
The present work is supported by the Korea Research Foundation Grant 
funded by the Korean Government(MOEHRD) (KRF-2006-312-C00507). The work of 
S.i.N. is partially supported by the  
Brain Korea 21 (BK21) project in Center of Excellency for Developing Physics 
Researchers of Pusan National University, Korea and by the grant for
Scientific Research (Priority Area No. 17070002) from the Ministry of
Education, Culture, Science and Technology, Japan.  The
authors thank the Department of Physics and Nuclear Physics \&
Radiation Technology Institute (NuRI), Pusan National University, where
this work was completed during the HNP07 on "Quarks in hadrons,
nuclei, and matter". S.i.N. would like to 
appreciate the fruitful comments from M.~Khlopov and Y.~Kwon.  

\end{document}